\documentclass[pre,aps,twocolumn,showpacs]{revtex4}
\begin{document}
\title{Solidity of viscous liquids. V. Long-wavelength dominance of the dynamics}
\author{Jeppe C. Dyre}
\affiliation{DNRF centre  ``Glass and Time,'' IMFUFA,  Department of Sciences, 
Roskilde University, Postbox 260, DK-4000 Roskilde, Denmark}
\date{\today}

\newcommand{\xik}{\xi_{\bf k}}\newcommand{\iomt}{i\omega\tau}\newcommand{\bu}{{\bf u}}\newcommand{\br}{{\bf r}}\newcommand{\bnul}{{\bf 0}}\newcommand{\bk}{{\bf k}}
\newcommand{\bom}{{\bf\Omega}}\newcommand{\bnabla}{{\bf\nabla}}\newcommand{\thn}{\theta_0}\newcommand{\la}{\left\langle}\newcommand{\ra}{\right\rangle}
\newcommand{\rk}[1]{{\rho_{\bf  {#1}}}}\newcommand{\fik}[1]{{\Phi_{\bf  {#1}}}(t)}\newcommand{\fnik}[1]{{\Phi^{(0)}_{\bf  {#1}}}(t)}\newcommand{\rok}{\rk\bk}
\newcommand{\romk}{\rk{-\bk}}\newcommand{\gr}{\Gamma_{\rho}}\newcommand{\sti}{\tilde s}\newcommand{\tti}{\tilde t}\newcommand{\mt}{\langle\rho\rangle}\newcommand{\fnul}{\Phi}

\begin{abstract}
This paper is the fifth in a series exploring the physical consequences of the solidity of highly viscous liquids. Paper IV proposed a model where the density field is described by a time-dependent Ginzburg-Landau equation of the nonconserved type with rates in $k$ space of the form $\Gamma_0+Dk^2$. If $a$ is the average intermolecular distance, the model assumes that $D\gg\Gamma_0a^2$. This inequality expresses a long-wavelength dominance of the dynamics, which implies that the Hamiltonian (free energy) to a good approximation may be taken to be ultralocal, i.e., with the property that equal-time field fluctuations are uncorrelated in space. Paper IV also briefly discussed how to generalize the model by including the molecular orientational fields, the stress tensor fields, and the potential energy density field. In the present paper it is argued that this is the simplest model consistent with the following three experimental facts: 1) Viscous liquids approaching the glass transition do not develop long-range order; 2) The glass has lower compressibility than the liquid; 3) The alpha process involves several decades of relaxation times shorter than the mean relaxation time. The paper proceeds to list six further experimental facts of viscous liquid dynamics and shows that these follow naturally  from the model. 
\end{abstract}

\pacs{64.70.Pf}

\maketitle

\section{Introduction}

The idea that viscous liquids approaching the calorimetric glass transition are solid-like goes back in time at least to Kauzmann's and Goldstein's famous papers from 1948 and 1969 \cite{kau48,gol69}. According to these authors, when a molecule changes position in a highly viscous liquid, this happens in the form of a so-called flow event, a sudden rearrangement of a group of molecules. In this picture, which was later confirmed by computer simulations \cite{heu97,sas98,ang00,sch00,sci05}, all motion is vibrational in the time between flow events. This indicates that a viscous liquid is much like a disordered solid \cite{note_flow}. 

The property that highly viscous liquids are solid-like and more to be viewed as ``solids that flow'' than like ordinary less-viscous liquids, is here termed solidity. This paper is the fifth in a series (I-IV) \cite{I,II,III,IV} devoted to extracting the physical consequences of the solidity of highly viscous liquids. A discussion of solidity, its motivation and consequences, may be found in the introduction to paper IV to which the reader is referred for further physical background.

A crucial ingredient of solidity is time-scale separation in the equilibrium viscous liquid. This is the fact that, whereas some processes occur on the vibrational time scale, i.e., over picoseconds, the genuine relaxation processes are much slower. Depending on temperature the latter occur on time scales of milliseconds, second, days,... The below discussion focuses exclusively on modelling the relaxation processes.

One paper utilizing arguments from solid-state elasticity to viscous liquid dynamics preceded this series. This was a joint publication with Olsen and Christensen from 1996 where the ``shoving model'' for the temperature dependence of the viscosity (or relaxation time) was proposed \cite{0}. According to this and related elastic models \cite{dyr06a,dyr06} the activation energy is proportional to the instantaneous shear modulus $G_\infty$ (the shear modulus measured on a very short time scale). This elastic constant increases significantly upon cooling, enough to explain the observed non-Arrhenius behavior for several molecular liquids \cite{0,dyr06,dyr06a} (more data, however, are needed to illuminate whether this generally explains the non-Arrhenius viscosity).

The first and second papers of this series focussed on the individual flow events. In paper I a ``solidity'' length $l$ was introduced characterizing the length scale below which there is time between two flow events to establish elastic equilibrium. In terms of the average intermolecular distance $a$, the alpha relaxation time $\tau$, and the high-frequency sound velocity $c_\infty$, the solidity length is given by $l^4=a^3\tau c_\infty$. Close to the calorimetric glass transition the solidity length approaches $1\,\mu\rm m$, so glass-forming liquids are solid-like on quite large length scales. The model discussed in paper IV and here focuses on dynamics below the solidity length.

Papers III and IV \cite{III,IV} dealt with the alpha relaxation process and how to explain its seemingly generic high-frequency loss $\propto\omega^{-1/2}$ \cite{ols01}. Paper III approached the problem {\it inductively} by noting that the BEL model from 1967 \cite{bar67} fits data for the frequency-dependent shear modulus well. Starting from the BEL model it was argued that the $\omega^{-1/2}$ high-frequency behavior arises from a long-time-tail mechanism operating over a range of times {\it shorter} than the alpha relaxation time. This was justified by a solidity-based argument with the irrelevance of momentum conservation as an important ingredient, thus allowing for the centre of mass to move following a flow event. As detailed in Refs. \cite{dyr07a,dyr07b} momentum conservation is irrelevant for highly viscous liquids, just as it is irrelevant in theories for defect motion in crystals.

Paper IV took a {\it deductive} approach and proposed a field-theoretic model giving a concrete realization of the idea of paper III. In this model, which is the subject also of the present paper, the generic $\omega^{-1/2}$ high-frequency decay of the alpha loss derives from a third-order term in the Hamiltonian (free energy) \cite{dyr05}. The model, besides regarding momentum conservation as irrelevant just as in other stochastic models (e.g., for polymer dynamics \cite{doi86}), was based on the observation that density conservation is also apparently disobeyed. This follows from solidity: A flow event brings the liquid from one potential energy minimum to another. Any potential energy minimum corresponds to a state of elastic equilibrium, thus a state of zero divergence of the stress tensor -- a ``solid'' state. From the standard equations of elasticity \cite{lan70} it follows that upon a local density change, the leading term in the displacement field in the surroundings is radial and varies with distance $r$ from the flow event as $1/r^2$. This is a pure shear displacement \cite{lan70}, implying that in a coarse-grained description density can change at one point without changing elsewhere; density has the appearance of not being conserved. This is conveniently condensed into the following equation for the coarse-grained density dynamics (where $b_\mu$ is a dimensionless measure of the magnitude of the flow event taking place at $\br_\mu$ at time $t_\mu$):

\begin{equation}\label{0}
\dot\rho(\br,t)\,=\,\sum_\mu b_\mu\delta(\br-\br_\mu)\delta(t-t_\mu)\,.
\end{equation}
Equation (\ref{0}) does not constitute a {\it theory}, of course, since it does not describe how flow events correlate; it just {\it describes} the equilibrium coarse-grained density fluctuations. Nevertheless, Eq. (\ref{0}) serves to emphasize that density has the appearance of not being conserved, a result which is less trivial than the already-mentioned momentum nonconservation deriving directly from the extremely large kinematic viscosity \cite{note_visc} characterizing liquids approaching the glass transition \cite{dyr07b}. The model discussed in paper IV and below incorporates the main idea behind Eq. (\ref{0}), density nonconservation, into a framework that is explicitly consistent with statistical mechanics. The model description of the density dynamics is more involved than that of Eq. (\ref{0}), which is too simple because it corresponds to isotropic flow events. On very large length scales, though, the model dynamics are described by Eq. (\ref{0}).

In the present paper, supplementing the arguments of papers III and IV, we adopt a combined {\it inductive/deductive} approach. First, we list three experimental facts characterizing highly viscous liquids from which the model is arrived at as the simplest model consistent with these facts; this constitutes the {\it inductive} part of the paper (Secs. 2 and 3). In Sec. 4 we proceed to discuss six further experimental facts and their relation to the model. It is argued that most of these are consequences of the model whereas some, though not mathematical consequences, appear quite natural when viewed in light of the model. This constitutes the {\it deductive} part of the paper. Section 5 gives  a brief discussion.

\section{Three facts of viscous liquid dynamics}

This section lists three facts characterizing glass-forming liquids approaching the calorimetric glass transition.

\begin{itemize}
\item {\it {\bf Fact 1:} There is no long-range density-coupled order}\end{itemize}
A popular approach to understanding why the viscosity increases ten orders of magnitude for a temperature decrease of typically just 10-15{\%} is to assume that some sort of  long-ranged order gradually develops upon supercooling. According to several prominent models the dramatic relaxation-time increase is a consequence of the liquid approaching a critical point where the relaxation time becomes infinite (see, e.g., the excellent recent reviews by Tanaka \cite{tan05}, Tarjus and coworkers \cite{tar05}, and Lubchenko and Wolynes \cite{lub07}, and their references). Following the theory of critical phenomena it is assumed that there is a diverging correlation length at the critical point. There is no consensus on how to define the proposed diverging correlation length, though, i.e., which quantity develops long-ranged correlations. Numerous X-ray and neutron scattering experiment however show that whatever this hypothetical quantity may be, it does not couple to the density field: No increase of long-range density fluctuations is observed upon cooling \cite{GT_ref}.

\begin{itemize}
\item {\it {\bf Fact 2:} The glassy phase has lower compressibility than the liquid phase.}\end{itemize}
The glass transition is a falling-out-of-equilibrium taking place when the liquid relaxation time becomes longer than the characteristic laboratory time scale. Compliance-type linear-response quantities like specific heat, compressibility, and thermal expansion coefficient all decrease when going from the liquid to the glass \cite{GT_ref}. This is easy to understand, because if each compliance-type linear-response quantity has contributions from both the fast (vibrational) and the much slower (configurational) degrees of freedom, these linear-response quantities must decrease at the glass transition since below $T_g$ the configurational degrees freeze and cease to contribute.

\begin{itemize}
\item {\it {\bf Fact 3:} The alpha process is characterized by a distribution of relaxation times covering several decades of times shorter than the alpha relaxation time.}\end{itemize}
Dielectric relaxation experiments are often fitted by the frequency-dependent response function corresponding to the stretched exponential dipole time-autocorrelation function, $\exp[-(t/\tau)^\beta]$ \cite{DR_ref}. This reflects the fact that viscous liquids with few exceptions \cite{note_debye} do not have simple exponential time-autocorrelation functions. There is nothing magic about the stretched exponential; it gives a good single-parameter fit to data because it reproduces the observed loss peaks that are always asymmetric to the high-frequency side \cite{DR_ref}. The point to be made here is that if one rewrites any good fit to the observed autocorrelation function as a sum of exponentials, the distribution function must include several decades of relaxation times shorter than the main (alpha) relaxation time.

\section{The simplest model for equilibrium viscous liquid dynamics consistent with facts 1-3}

In this section we consider the question: Which variable(s) must be included in a useful theory?  What are the simplest dynamics for this/these variable(s) consistent with facts 1-3? Most models for viscous liquid dynamics attempt to explain both the alpha relaxation characteristics at a given temperature and the non-Arrhenius temperature dependence of the alpha relaxation time. We here leave aside the non-Arrhenius problem altogether, although solidity appears to play an important role for this property as well \cite{dyr06}. The idea is to ``cut the Gordian Knot'' by separating the complex problem of viscous liquid dynamics into two independent, hopefully easier problems.

Nowadays most of physics -- from particle physics to critical phenomena, electromagnetism, and condensed-matter physics in quite diverse  contexts -- is formulated in the language of field theory. It is natural to expect that viscous-liquid dynamics should also be described by a field theory \cite{dic07}. The question which variables are relevant thus becomes: ``Which fields must be included in the description?''

The obvious fields are those of standard hydrodynamics: the momentum, energy, and particle density fields. As mentioned, the conservation laws for momentum and energy are both irrelevant for viscous liquids \cite{dyr07a,dyr07b}. Thus the standard hydrodynamic description, based on continuity equations for these quantities, looses its physical significance. Despite the fact that solidity also implies (apparent) density nonconservation, the situation is different for the particle density field -- after all, molecules are not continuously exchanged with the surroundings in the way that momentum and energy are. We thus base the model sought for on the density field $\rho(\br,t)$. The question which other fields to include in order to have a complete description of the macroscopic dynamics, is dealt with at the end of this section.

The next question is: What are the simplest possible dynamics? Density dynamics of viscous liquids have two parts, the vibrations (phonons) and the ``relaxing'' part of the dynamics. At low temperatures the alpha relaxation time is much larger than picoseconds, and the two dynamics are very well separated. Thus it makes good sense to ignore the vibrational part of the dynamics. Assuming that the relevant field theory is based on a Hamiltonian (i.e., a free energy functional), the question next is how to model the dynamics consistent with the Hamiltonian. The answer to this is well-known, use Langevin dynamics that is {\it the} generic way to arrive at dynamics from statics \cite{doi86,STOCH_ref}: If the relevant variables are denoted by $Q_1,...,Q_n$, the Hamiltonian is $H(Q_1, ..., Q_n)$, and $\beta=1/k_BT$, the Langevin equation is $\dot Q_i=-\Gamma_i\partial_i(\beta H)/\partial Q_i+\xi_i(t)$ where $\xi_i(t)$ is a Gaussian white noise term with zero mean obeying $\langle\xi_i(t)\xi_j(t')\rangle=2\Gamma_i\delta_{ij}\delta(t-t')$. These equations reproduce the canonical probability $\propto\exp(-\beta H)$, thus ensuring consistency with statistical mechanics \cite{doi86,STOCH_ref}.

The system consists of $N$ molecules with coordinates $\br_j$ in volume $V$, and the density field is defined by $\rho(\br)=\sum_j\delta(\br-\br_j)$. As always when there is translational invariance it is convenient to go to k-space; the range of allowed k-vectors is limited to the discrete set consistent with periodic boundary conditions. We define the k'th density component as

\begin{equation}\label{1}
\rk\bk\,=\,
\frac{1}{\sqrt N}
\sum_j e^{i\bk\cdot\br_j}\,.
\end{equation}
With this normalization $\rk\bk$ fluctuations become independent of volume for $V\rightarrow\infty$ ($N/V=\rm Const.$) and the relaxational part of the static structure factor $S(k)$ is given by $S(k)=\langle\rk\bk\rk{-\bk}\rangle$ for $V\rightarrow\infty$. The Langevin equation \cite{doi86,STOCH_ref} is 

\begin{equation}\label{2}
\dot\rho_\bk\,=\,
-\Gamma_k \frac{\partial(\beta H)}{\partial \romk}\,+\,\xik(t)\,.
\end{equation}
The complex Gaussian white noise term obeys $\xik^*(t)=\xi_{-\bf k}(t)$ and $\langle\xik(t)\xi_{\bf k'}^*(t')\rangle=2\Gamma_k\delta_{\bf k,k'}\delta(t-t')$. Equation (\ref{2}) is a standard time-dependent Ginzburg-Landau equation. Because $\rho_{\bf k}^*=\rho_{-\bf k}$ this equation in conjunction with the equation for $\rho_{-\bf k}$ is equivalent to two real Langevin equations, one for the real part of $\rho_{\bf k}$ and one for its imaginary part. 

Following standard field theory procedure we split $H$ into a sum of a quadratic ``free-field'' term $H_0$ and an ``interaction'' term $H'$ containing all higher-order terms:

\begin{equation}\label{3}
H\,=\,H_0+H'\,.
\end{equation}
Regarding $H'$ as a perturbation, let us focus on the dynamics embodied in the free-field time-autocorrelation function denoted by $\langle\rk\bk(0)\rk{-\bk}(t)\rangle_0$. If there were no higher order terms, because $S(k)=\langle\rk\bk\rk{-\bk}\rangle$, the free-field Hamiltonian would be given by

\begin{equation}\label{4}
\beta H_0\,=\,
\frac{1}{2}
\sum_\bk\frac{\rk\bk\rk{-\bk}}{S(k)}\,.
\end{equation}
Substituting this into Eq. (\ref{2}) we find that the free-field time-autocorrelation function $\langle\rk\bk(0)\rk{-\bk}(t)\rangle_0$ is an exponential with decay rate $\gamma_k$ given by 

\begin{equation}\label{5}
\gamma_k\,=\,\frac{\Gamma_k}{S(k)}\,.
\end{equation}
When $H'$ is just a perturbation, the distribution of relaxation rates is roughly given by the distribution of $\gamma_k$'s. 

Since molecules cannot disappear, one would {\it a priori} assume $\Gamma_k\propto k^2$ (``conserved'' case -- model B of Ref. \cite{hoh77}), reflecting the expectation that density at sufficiently long wavelengths obeys the diffusion equation. This, however, is inconsistent with fact 2 for the following reason. At the glass transition the liquid high-frequency compressibility becomes the glass compressibility. Fact 2 states that this quantity is lower than the liquid (dc) compressibility, implying that in the equilibrium liquid there are relaxational volume fluctuations on a macroscopic length scale taking place on a finite time scale (the alpha time scale). (This is also known from measurements of the frequency-dependent bulk modulus showing that the low-frequency bulk modulus is smaller than the high-frequency bulk modulus \cite{chr94}, as well as from the so-called Mountain peak (relaxation mode) of light scattering \cite{mou68,ber76}.) This implies that the relaxational part of the static structure factor, $S(k)$, has a non-zero limit for $k\rightarrow 0$. Therefore, if $\Gamma_k\propto k^2$, by Eq. (\ref{5}) the rate of the relaxational macroscopic density fluctuations would go to zero for $k\rightarrow 0$, i.e., be arbitrarily slow for large enough samples. This violates fact 2. Consequently, density fluctuations cannot be described by $\Gamma_k\propto k^2$ as $k\rightarrow 0$, and density must have the appearance of a {\it non-conserved} field.

Given that $\Gamma_k\propto k^2$ does not work, the simplest alternative is that  $\Gamma_k$ is independent of k: $\Gamma_k=\Gamma_0$ (``non-conserved'' case -- model A of Ref. \cite{hoh77}). This, however, does not work for the following reason. For $ka\sim 1$ the static (relaxational) structure factor $S(k)$ is of order one, whereas for $k\rightarrow 0$ $S(k)$ converges to the ratio between liquid (relaxational) compressibility and that of an ideal gas at the same density. This ratio is typically of order $10^{-2}$ \cite{LIQ_ref}. Thus if $\Gamma_k=\Gamma_0$, Eq. (\ref{5}) would imply a range of relaxation times covering at most 2 decades. This is inconsistent with fact 3 that requires several decades of relaxation times. (It would also be physically counterintuitive to have long-wavelength density fluctuations decaying much faster than short-wavelength fluctuations.)

Since $\Gamma_k\propto k^2$ contradicts fact 2 and $\Gamma_k=\Gamma_0$ contradicts fact 3, the next possibility is a combination of the two:

\begin{equation}\label{6}
\Gamma_k\,=\,\Gamma_0+Dk^2\,.
\end{equation}
For this to be consistent with fact 3, however, $D$ must be quite large: Unless the $Dk^2$ term makes $\Gamma_k$ vary several decades for the range of allowed k-vectors, the model won't work for the same reason that $\Gamma_k=\Gamma_0$ doesn't work. Since the maximum k obeys $ka\sim 1$, this means that the following must be assumed:

\begin{equation}\label{7}
D\,\gg\,\Gamma_0a^2\,.
\end{equation}
The inequality (\ref{7}) expresses a {\it long-wavelength dominance of the dynamics} because it implies that the $Dk^2$ term dominates the rate expression Eq. (\ref{6}) for a range of small k vectors corresponding to wavelengths much larger than $a$. Whereas in papers III and IV this inequality was justified by microscopic arguments, it here comes about from a search for the simplest possible model consistent with facts 1-3. 

In conjunction with fact 1, long-wavelength dominance of the dynamics implies that static (equal-time) correlations of density fluctuations at differing points can have little influence on the dynamics. Consequently, in the name of simplicity such correlations will be assumed to be absent altogether. A field theory with no equal-time correlations between fields at different points in space is termed ``ultralocal.'' In k-space the assumption of ultralocality means that all coefficients are k independent in the expansion in orders of $\rk\bk$: $H_0=\sum_\bk(1/2A) \rk\bk\rk{-\bk}$ and $H'=\sum_{\bk,\bk'}(\lambda_3/3\sqrt N)\rk\bk \rk {\bk'} \rk{-\bk-\bk'}+...$ \cite{note_thermo}.

A similar long-wavelength dominance of the deeply supercooled dynamics is crucial also in Schweizer's and Saltzman's theory of activated hopping in polymer melts \cite{sch_papers}. Specifically, atomistic structure in S(k) is here coarse-grained over, resulting in a statistical segment level (nm-scale) description where confining forces are quantified via the amplitude of thermal density fluctuations, which are proportional to the isothermal compressibility, i.e., given by S(0). In other words, it is assumed that the most important dynamic length scale is significantly larger than a monomer.

The assumption that equal-time density fluctuations are uncorrelated in space does not mean that the model is trivial -- there are non-zero correlations between density fluctuations at different positions {\it at different times}. The situation is analogous to that of a trivial spin model (i.e.,  with no spin-spin interactions) with Kawasaki dynamics, where a spin flip at one point can take place only if a neighboring spin simultaneously flips in the opposite direction.  If by chance a given up spin is surrounded by a large island of up spins, the given spin will be frozen for some time; thus the dynamics of this spin is influenced by those of its surrounding spins even though there are no equal-time spin-spin correlations. 

As a minimum, a theory for viscous liquid dynamics should make it possible to calculate all macroscopic frequency-dependent linear-response quantities. Response functions are determined from the fluctuation-dissipation theorem by first calculating time-autocorrelation functions of variables like the total dipole moment (dielectric constant), pressure (bulk modulus), shear stress (shear modulus), or energy density (specific heat). Note that the frequency-dependent bulk modulus may be determined from density fluctuations alone, implying that the pressure is not an independent variable in the present context, and the stress tensor has only 5 relevant components. 

These considerations lead (paper IV) to the following general recipe for modelling viscous liquid dynamics:
\begin{enumerate}
\item The relevant degrees of freedom are fields $\phi^{(1)}(\br)$, ...,  $\phi^{(n)}(\br)$ defined as: a) the densities of the different molecules, b) the densities of the molecules' various configurational variables reflecting the molecular symmetry, c) the 5 stress tensor fields of the traceless stress tensor, d) the potential energy density field;
\item The Hamiltonian $H$ (free energy) is ultralocal and consists of scalar (i.e., rotationally and translationally invariant) terms; 
\item For each field the dynamics are described by a time-dependent Ginzburg-Landau equation, i.e., a Langevin equation of the form

\begin{equation}
\dot \phi^{(j)}_\bk=-\Gamma^{(j)}_k\,\frac{\partial (\beta H)}{\partial \phi^{(j)}_{-\bk}}+\xi^{(j)}_\bk(t)\,, 
\end{equation}
where $\xi^{(j)}_\bk(t)$ is a Gaussian white-noise term obeying  
$\langle \xi^{(i)}_\bk(t) \xi^{(j)*}_{\bk '}(t)\rangle=2\Gamma^{(i)}_k\delta_{i,j}\delta_{\bk,\bk '}\delta(t-t')$;
\item For each density field the Langevin equation rates are given by $\Gamma^{(j)}_k=\Gamma^{(j)}_0+D^{(j)} k^2$ where $D^{(j)}\gg\Gamma^{(j)}_0 a^2$; for all other fields the rates are k-independent: $\Gamma^{(j)}_k=\Gamma^{(j)}_0$.
\end{enumerate}
Inclusion of the extra fields makes it possible to calculate the frequency-dependent dielectric constant and shear modulus, as well as all 8 fundamental frequency-dependent thermoviscoelastic response functions \cite{ell07}: the isochoric and isobaric specific heats, the isothermal and adiabatic compressibilities, the isobaric expansion coefficient, the adiabatic contraction coefficient, and the isochoric and adiabatic pressure coefficients.

\section{Six further facts of viscous liquid dynamics and their interpretation in light of the model}

Below are listed six experimental facts that are each nontrivial in the sense that there is no logically compelling connection between it and facts 1-3. Nevertheless, when viewed in light of the model, these facts all appear obvious.

\begin{itemize}
\item {\it {\bf Fact 4:} Below the alpha-loss peak the loss virtually follows the Debye prediction, i.e., is proportional to $\omega$. Thus there is an effective cut-off at long times in the relaxation-time distribution.}\end{itemize}
Alpha loss peaks, which are typically measured by dielectric relaxation experiments, are generally asymmetric towards the high-frequency side \cite{DR_ref}. At low frequencies they follow the Debye prediction [$\propto 1/(1+\iomt)$], i.e., the loss is virtually proportional to frequency; the same applies, e.g., for the frequency-dependent shear modulus \cite{jak05}. This implies that, if the linear response is written formally as a sum of Debye processes, there is an effective cut-off at long times in the relaxation-time distribution.

Given the spatial disorder of any viscous liquid it is not surprising that Debye peaks are rarely observed, but the observed effective cut-off at long relaxation times is highly nontrivial. If the non-Debye relaxation were due to effects of disorder-induced activation-energy broadening, the obvious guess would be a Gaussian activation-energy distribution. This, however, implies loss peaks that are symmetric in a log-log plot, which is inconsistent with experiment.

Because there is a minimum relaxation rate in the model, a long relaxation-time cut-off is automatically implied. Of course, this cut-off was built into the model via Eq. (\ref{6}). The point we wish make is merely that this equation was not justified from fact 4, but as the simplest way to rationalize facts 1-3. Fact 4 follows.

\begin{itemize}
\item {\it {\bf Fact 5:} Above the alpha loss-peak frequency the loss appears to be generically close to $\omega^{-1/2}$.}\end{itemize}
The alpha process is conveniently monitored by dielectric relaxation experiments \cite{DR_ref}. Above the loss-peak frequency the dielectric loss decays following an approximate power law $\propto\omega^{-\nu}$. The shape of the loss peak often changes with temperature, i.e., time-temperature superposition (TTS) is often not obeyed. It is now generally agreed that Johari-Goldstein beta processes may be found at much lower frequencies than previously thought \cite{ols98}. Thus, since alpha and beta relaxations have quite different temperature dependence, low-lying beta processes easily lead to TTS violations even in the Hertz regime and below. According to this reasoning the ``generic'' characteristics of the alpha process is observed only when TTS is accurately obeyed. A study of the dielectric loss of 10 molecular liquids published in 2001 \cite{ols01} indicated that when TTS is accurately obeyed, the exponent $\nu$ is quite close to $1/2$. A recent study comprising 45 molecular liquids found that the minimum slope $\alpha_{\rm min}$ above the loss peak (characterizing the inflection point and thus giving the best approximate power-law fit) is generally fairly close to $-1/2$ (60\% of the liquids studied obey $-0.6<\alpha_{\rm min}<-0.4$) \cite{albena}. 

As shown in Ref. \cite{dyr05} and paper IV, to lowest order in perturbation theory a third-order term in the Hamiltonian implies that the loss varies as $\omega^{-1/2}$ for $\omega$'s considerably higher than the loss peak frequency.

\begin{itemize}
\item {\it {\bf Fact 6:} The alpha relaxation process is dominated by small-angle jumps.}\end{itemize}
Many models assume ``cooperatively rearranging regions'' that do not interact with one another. In this spirit molecules either move significantly (those involved in the flow event) or do not move at all (those in the surroundings). Thus molecular jumps would be expected to be fairly large and molecular orientations likewise to change considerably following a flow event. The molecular jump angles cannot be probed by linear-response experiments, but fortunately they can by NMR experiments. The result of B{\"o}hmer and coworkers \cite{boh98,boh01} is that small-angle jumps dominate.

This observation calls for an explanation in terms of solidity: The reason that small-angle jumps dominate must be that -- if the flow event picture is not completely wrong -- the overall picture is dominated by the small adjustments in the surroundings required to re-establish elastic equilibrium after a flow event. A simple solidity-based calculation (paper I) shows that the jump-angle distribution $P(\phi)$ varies as $1/\phi^2$, consistent with NMR findings \cite{boh98} (this distribution is not normalizable because there are infinitely many molecules in the surroundings -- in reality the distribution is cut-off at very small angles because elastic effects do not propagate beyond the solidity length). Clearly, largest weight is given to small jump angles.

When fact 6 is contemplated in light of the model, it should be noted (paper IV) that the crucial $\Gamma_0\neq 0$ identity expressing apparent density nonconservation can come about only if a flow event is followed by small solidity-based adjustments of molecular positions in the far surroundings. Thus fact 6 follows from the model's nonzero $\Gamma_0$.

\begin{itemize}
\item {\it {\bf Fact 7:} Viscous slowing down is not accompanied by significant changes of the static structure factor.}\end{itemize}
A popular and obvious explanation of the dramatically increasing relaxation time for liquids approaching the calorimetric glass transition is that upon cooling there is a gradual build-up of some sort of long-range order, in many models signalling that there is a critical point not far below $T_g$ where the relaxation time becomes infinite \cite{tan05,tar05,lub07}. As mentioned, numerous experiments have looked for long-range order, but found none \cite{GT_ref}. Not only is there no long-range order, but the liquid structure as probed by $S(k)$ via X-ray or neutron scattering experiments changes little over the temperature range where the relaxation time changes by 10 or more orders of magnitude. It is matter of taste whether or not one regards this as surprising \cite{note_mc}. A model with no spatial correlations at any temperature trivially predicts that there are no nontrivial changes of the static structure factor upon cooling. In this sense fact 7 follows from the model.

\begin{itemize}
\item {\it {\bf Fact 8:} The Debye-Stokes-Einstein relation is often violated in viscous liquids.}\end{itemize}
An important  finding of the 1990's was that translational motions often decouple from and become 1-3 decades faster than rotations \cite{violations}. Somehow, translations are enhanced compared to what one expects from the Debye-Stokes-Einstein relation that estimates the single particle diffusion constant $D_s$ from the viscosity $\eta$. Although this relation {\it a priori} applies only for macroscopic particles, there is no obvious reason that a molecule on average should move much longer than $a$ during the rotational correlation time (the alpha relaxation time).

The generally accepted picture of Debye-Stokes-Einstein violations is that these reflect dynamic heterogeneity \cite{violations}. Although dynamic heterogeneity is not described by the model, this is not inconsistent with the model that provides only a coarse-grained description of the dynamics. Debye-Stokes-Einstein violations, in fact, fit nicely with the model if one assumes that the single-particle diffusion constant is roughly equal to the diffusion constant $D$ of Eq. (\ref{6}); in this case the long-wavelength dominance inequality (\ref{7}) simply expresses Debye-Stokes-Einstein violation.

\begin{itemize}
\item {\it {\bf Fact 9:} The dynamics are not sensitive to the chemistry.}\end{itemize}
At any given temperature the relaxation characteristics -- linear as well as nonlinear -- are similar for all glass-forming liquids. Viscous liquids do differ as regards, e.g., whether or not there is a clearly defined beta relaxation in the liquid phase, or whether or not time-temperature superposition applies. But these differences do not appear to correlate with chemical characteristics in any simple fashion. For instance, as regards their dielectric and thermodynamics properties, liquids held together exclusively by van der Waals forces do not differ systematically from those involving hydrogen bonds \cite{DR_ref}. Similarly, the frequency-dependent shear modulus of viscous metallic liquids forming bulk metallic glasses is indistinguishable from that of typical molecular liquids \cite{china_acad}. The overall conclusion is that it is not possible from macroscopic measurements alone to determine which chemical bonds are involved (except, of course, from the fact that the temperature range where the glass transition takes place trivially provides information about the strength of the intermolecular forces). As a further confirmation of fact 9 it should be noted that mixtures behave much like pure liquids and that, again, it is not generally possible from purely macroscopic measurements to tell whether or not a given liquid is a mixture.

This insensitivity to the chemistry is a highly significant fact. For some reason the dynamics are fairly indifferent to details of the molecular interactions, but why? In the model chemistry independence is expected because of the long-wavelength dominance of the dynamics -- clearly chemistry plays little role for dynamics on length scales much larger than the size of a molecule. The chemistry independence is similar to that of critical phenomena. This has motivated many attempts to draw parallels to the theory of critical phenomena by assuming that viscous slowing down is accompanied by some kind of long-range order responsible for the observed quasi-universality. The present model has no such assumption, but on the contrary assumes that there are only {\it short-range} static correlations of the relevant fields. Only when it comes to the {\it dynamics} do long length scales play important roles. -- Note that the model does not imply absolute chemistry independence, because the free energy as function of the fields varies with chemical composition.

\section{Discussion}

It is a long-standing assumption that viscous-liquid dynamics are cooperative. Mode-coupling theory \cite{got92,das04} is an interesting case where cooperativity enters via the coupling of single-particle motion to the surroundings, resulting in a modification of the single-particle motion with drastic consequences at low temperatures (infinite relaxation time at a finite temperature in the simplest and most studied version of the theory). The present model is also cooperative, but in a rather simple-minded sense, with two elements of cooperativity: 1) Density nonconservation (implying $\Gamma_0\neq 0$)  is a cooperative effect because it involves the solidity-based movement of molecules far from a flow event; 2) The long-wavelength assumption Eq. (\ref{7}) expresses cooperativity in the sense that motion over long wavelengths is mainly responsible for the alpha process.

The overall purpose of this paper was to show that the model of paper IV has implications that were not put into the model. The overall credibility of the model is strengthened by the fact that the model is consistent with -- and in most cases predicts -- several experimental facts that are independent of the model inputs. We would like finally to emphasize that, despite the macroscopic reasoning of this paper, the long-wavelength dominance inequality Eq. (\ref{7}) reflects properties of the individual flow events \cite{note_flow_event}.

\acknowledgments 
The author wishes to thank Ken Schweizer for most useful correspondences and discussions on the model and its underlying assumptions. This work was supported by a grant from the Danish National Research Foundation (DNRF) for funding the centre for viscous liquid dynamics ``Glass and Time.''


\begin{thebibliography}{99}

\bibitem{kau48} W. Kauzmann, Chem. Rev. {\bf 43}, 219 (1948).

\bibitem{gol69} M. Goldstein, J. Chem. Phys. {\bf 51}, 3728 (1969).

\bibitem{heu97} A. Heuer, Phys. Rev. Lett. {\bf 78}, 4051 (1997).

\bibitem{sas98} S. Sastry, P. G. Debenedetti, and F. H. Stillinger, Nature {\bf 393}, 554 (1998).

\bibitem{ang00} L. Angelani, R. Di Leonardo, G. Ruocco, A. Scala, and F. Sciortino, Phys. Rev. Lett.  {\bf 85}, 5356 (2000).

\bibitem{sch00} T. B. Schr{\o}der, S. Sastry, J. C. Dyre, and S. C. Glotzer, J. Chem. Phys. {\bf 112}, 9834 (2000).

\bibitem{sci05} F. Sciortino, J. Stat. Mech., Art. No. P05015 (2005). 

\bibitem{note_flow} Solids also flow when subjected to an external stress. The primary difference between a disordered solid and a highly viscous liquid is that the latter is in a state of metastable equilibrium, i.e., fully characterized by two thermodynamic variables (e.g., temperature and pressure). Thus the highly viscous liquid and its dynamics ought to be much simpler.

\bibitem{I} J. C. Dyre, Phys. Rev. E {\bf 59}, 2458 (1999) (paper I).

\bibitem{II}  J. C. Dyre, Phys. Rev. E {\bf 59}, 7243 (1999) (paper II).

\bibitem{III}  J. C. Dyre, Phys. Rev. E {\bf 72}, 011501 (2005) (paper III).

\bibitem{IV}  J. C. Dyre, Phys. Rev. E {\bf 74}, 021502 (2006) (paper IV).

\bibitem{0} J. C. Dyre, N. B. Olsen, and T. Christensen, Phys. Rev. B {\bf 53}, 2171 (1996). 

\bibitem{dyr06a} J. C. Dyre, T. Christensen, and N. B. Olsen, J. Non-Cryst. Solids {\bf 351}, 4635 (2006).

\bibitem{dyr06} J. C. Dyre, Rev. Mod. Phys. {\bf 78}, 953 (2006).

\bibitem{ols01} N. B. Olsen, T. Christensen, and J. C. Dyre, Phys. Rev. Lett. {\bf 86}, 1271 (2001).

\bibitem{bar67} A. J. Barlow, A. Erginsav, and J. Lamb, Proc. Roy. Soc. London, Ser. A {\bf  298}, 481 (1967).

\bibitem{dyr07a} J. C. Dyre, J. Phys.: Condens. Matter {\bf 19}, 205105 (2007).

\bibitem{dyr07b} J. C. Dyre, Philos. Mag. {\bf 87}, 497 (2007).

\bibitem{dyr05} J. C. Dyre, Europhys. Lett. {\bf 71}, 646 (2005).

\bibitem{doi86} M. Doi and S. F. Edwards, {\it Theory of Polymer Dynamics} (Oxford University Press, Oxford, 1986).

\bibitem{lan70} L. D. Landau and E. M. Lifshitz, {\it Theory of Elasticity}, 2nd Ed. (Pergamon, London, 1970).

\bibitem{note_visc} By an unfortunate mistake, in Ref. \cite{dyr06} the ``kinematic viscosity'' of the Navier-Stokes equation (viscosity over density, a quantity with same dimension as a diffusion constant) was referred to as the ``dynamic viscosity.'' The kinematic viscosity is the transverse momentum diffusion constant, a quantity that becomes very large when a liquid approaches the calorimetric glass transition. This is responsible for the fact that momentum is exchanged between liquid and measuring equipment over the alpha time scale, thus implying irrelevance of momentum conservation for realistic descriptions of viscous liquid dynamics as it is observed in the laboratory.

\bibitem{tan05} H. Tanaka, J. Non-Cryst. Solids {\bf 351}, 3371 (2005); {\bf 351}, 3385 (2005); {\bf 351}, 3396 (2005).

\bibitem{tar05} G. Tarjus, S. A. Kivelson, Z. Nussinov,  and P. Viot, J. Phys.: Condens. Matter {\bf 17},  R1143 (2005).

\bibitem{lub07} V. Lubchenko and P. G. Wolynes, Ann. Rev. Phys. Chem. {\bf 58}, 235 (2007). 

\bibitem{GT_ref} 
G. Harrison, {\it The dynamic properties of supercooled liquids} (Academic Press, New York, 1976);
J. Lamb, J. Rheol. {\bf 22}, 317 (1978);
S. Brawer, {\it Relaxation in Viscous Liquids and Glasses} (American Ceramic Society, Columbus, OH, 1985);
I. M. Hodge, J. Non-Cryst. Solids {\bf 169}, 211 (1994);
C. A. Angell, K. L. Ngai, G. B. McKenna, P. F. McMillan, and S. W. Martin, J. Appl. Phys. {\bf 88}, 3113 (2000);
C. Alba-Simionesco, C. R. Acad. Sci. Paris (Ser. IV) {\bf 2}, 203 (2001);
P. G. Debenedetti and F. H. Stillinger, Nature {\bf 410}, 259 (2001);
E. Donth, {\it The Glass Transition} (Springer, Berlin, 2001);
U. Buchenau and A. Wischnewski, Phys. Rev. B {\bf 70}, 092201 (2004);
W. Kob, in {\it Slow relaxations and nonequilibrium dynamics in condensed matter, Proceedings of the Les Houches Summer School of Theoretical Physics, Session LXXVII, 1-26 July, 2002}, p. 199, edited by J.-L. Barrat, M. Feigelman, J. Kurchan, and J. Dalibard (Springer, Berlin, 2004); 
G. N. Greaves and S. Sen, Adv. Phys. {\bf 56}, 1 (2007).

\bibitem{DR_ref}
A. Kudlik, S. Benkhof, T. Blochowicz, C. Tschirwitz, and E. R{\"o}ssler, J. Mol. Structure {\bf 479}, 201 (1999);
P. Lunkenheimer, U. Schneider, R. Brand, and A. Loidl, Contemp. Phys. {\bf 41}, 15 (2000);
F. Kremer and A. Sch{\"o}nhals (Eds.), {\it Broadband Dielectric Spectroscopy} (Springer, Berlin, 2002);
C. M. Roland, S. Hensel-Bielowka, M. Paluch, R. Casalini, Rep. Prog. Phys. {\bf 68}, 1405 (2005).

\bibitem{note_debye} There is one interesting exception, namely the lowest relaxation process of mono-hydroxy-alcohols that is purely Debye, a process which is much slower than the ordinary alpha process determining the calorimetric $T_g$, see, e.g., L. M. Wang, S. Shahriari, and R. Richert, J. Phys. Chem. B {\bf 109}, 23255 (2005).

\bibitem{dic07} D. di Caprio and J. P. Badiali, arXiv:0707.3069 (2007).

\bibitem{STOCH_ref} N. G. van Kampen, {\it Stochastic Processes in Physics and Chemistry} (North-Holland, Amsterdam, 1981);
H. Risken, {\it The Fokker-Planck Equation}, 2nd Ed. (Springer, Berlin, 1989);
L. E. Reichl, {\it  A Modern Course in Statistical Physics}, 2nd Ed. (Wiley, New York, 1998).

\bibitem{hoh77} P. C. Hohenberg and B. I. Halperin, Rev. Mod. Phys. {\bf 49}, 435 (1977).

\bibitem{chr94} T. Christensen and N. B. Olsen, Phys. Rev. B {\bf 49}, 15396 (1994).

\bibitem{mou68} R. D. Mountain, J. Res. Natl. Bur. Stand. {\bf 72A}, 95 (1968).

\bibitem{ber76} B. J. Berne and R. Pecora, {\it Dynamic Light Scattering} (Wiley, New York, 1976 - Dover edition 2000).

\bibitem{LIQ_ref} J. P. Boon and S. Yip, {\it Molecular hydrodynamics} (McGraw-Hill, New York, 1980);
J. P. Hansen and I. R. McDonald, {\it Theory of simple liquids} 2nd Ed. (Academic Press, New York, 1986);
J.--L. Barrat and J.--P. Hansen, {\it Basic concepts for simple and complex liquids} (Cambridge University Press, Cambridge, 2003).

\bibitem{note_thermo} Note that the assumption of an ultralocal Hamiltonian means that there is a direct connection to the macroscopic thermodynamics of the relaxing degrees of freedom (i.e., the non-vibrational contribution to the thermodynamics), because the Hamiltonian is nothing but the free energy of the relaxing degrees of freedom.

\bibitem{sch_papers} K. S. Schweizer and E. J. Saltzman, J. Chem. Phys. {\bf 121}, 1984 (2004);  E. J. Saltzman and K. S. Schweizer, J. Phys.: Condens. Matter {\bf 19}, 205123 (2007). 

\bibitem{ell07} N. L. Ellegaard, T. Christensen, P. V. Christiansen, N. B. Olsen, U. R. Pedersen, T. B. Schr{\o}der, and J. C. Dyre, J. Chem. Phys. {\bf 126}, 074502 (2007).

\bibitem{jak05} B. Jakobsen, K. Niss, and N. B. Olsen, J. Chem. Phys. {\bf 123}, Art. No. 234511 (2005).

\bibitem{ols98} N. B. Olsen, J. Non-Cryst. Solids {\bf 235}, 399 (1998).

\bibitem{albena} A. I. Nielsen et al., in preparation (2007).

\bibitem{boh98} R. B{\"o}hmer and G. Hinze, J. Chem. Phys. {\bf 109}, 241 (1998).

\bibitem{boh01} R. B{\"o}hmer, G. Diezemann, G. Hinze, and E. R{\"o}ssler, Prog. Nucl. Magn. Reson. Spectrosc. {\bf 39}, 191 (2001).

\bibitem{note_mc} Mode-coupling theory, of course, is based on the assumption that the dramatic slowing down is {\it caused} by the quite small structural changes that are indeed observed upon cooling. This fact makes mode-coupling theory in its classical form less compelling. Schweizer and Salzman developed a promising version of mode-coupling theory, which allows for barrier transitions and, via well-defined approximations, for calculating the alpha relaxation energy barrier from first principles [K. S. Schweizer and E. J. Salzman, J. Chem. Phys. {\bf 119}, 1181 (2003); K. S. Schweizer, J. Chem. Phys. {\bf 123}, 244501 (2005), E. J. Salzman and K. S. Schweizer, J. Chem. Phys. {\bf 125}, 044509 (2006)]. Very recently Schweizer showed how this novel approach connects to a number of other theories [K. S. Schweizer, J. Chem. Phys., submitted May 2007].

\bibitem{violations} F. Fujara, B. Geil, H. Sillescu, and G. Fleischer, Z. Phys. B {\bf 88}, 195 (1992);
F. H. Stillinger and J. A. Hodgdon, Phys. Rev. E {\bf 50}, 2064 (1994);
M. T. Cicerone and M. D. Ediger, J. Chem. Phys. {\bf 104}, 7210 (1996); 
G. Diezemann, H. Sillescu, G. Hinze, and R. B{\"o}hmer, Phys. Rev. E {\bf 57}, 4398 (1998); 
J. F. Douglas and D. Leporini, J. Non-Cryst. Solids {\bf 235-237}, 137 (1998);
S. F. Swallen, P. A. Bonvallet, R. J. McMahon, and M. D. Ediger, Phys. Rev. Lett. {\bf 90}, 015901 (2003).

\bibitem{china_acad} P. Wen, D. Q. Zhao, M. X. Pan, W. H. Wang, Y. P. Huang, and M. L. Guo, Appl. Phys. Lett. {\bf 84}, 2790 (2004); 
Z. F. Zhao, P. Wen, C. H. Shek, and W. H. Wang, Phys. Rev. B {\bf 75}, 174201 (2007).

\bibitem{got92} W. G{\"o}tze and L. Sj{\"o}gren, Rep. Progr. Phys.  {\bf 55}, 241 (1992).

\bibitem{das04} S. P. Das, Rev. Mod. Phys. {\bf 76}, 785 (2004).

\bibitem{note_flow_event} On a very short time scale flow events are uncorrelated and thus the inequality Eq. (\ref{7}) translates directly into properties of single flow events: If one flow event changes $\rk\bk$ by $\delta\rk\bk$, one finds from paper IV that $\langle|\delta\rk\bk|^2\rangle=\langle b^2\rangle/N$ for $k\rightarrow 0$ where $b$ (compare Eq. (\ref{0})) is defined by the radial displacement far from the flow event varying as $-b/(4\pi\langle\rho\rangle)1/r^2$ (paper IV), whereas the term $\propto k^2$ is$k^2\langle{\Delta\bf R}^2\rangle/(3N)$ where ${\Delta\bf R}$ is the vector sum of all molecule displacements induced by the flow event. On the other hand, integrating the noise term over a short time shows that $\langle|\delta\rk\bk|^2\rangle\propto \Gamma_0+Dk^2$. Comparing these expressions shows that the inequality Eq. (\ref{7}) is equivalent to $\langle b^2\rangle\ll \langle{\Delta\bf R}^2\rangle/a^2$. This could be obeyed either if $|b|\ll 1$ and $\langle{\Delta\bf R}^2\rangle\sim a^2$, or if $|b|\sim 1$ and the sum of all molecule displacements is much larger than the intermolecular distance $a$.

\end{thebibliography}
\end{document}